\documentclass[prb,aps,showpacs,twocolumn,unsortedaddress,superscriptaddress]{revtex4-1}
\usepackage{graphics, bm, psfrag, amsmath, amssymb, epsfig, grffile, float, bbold}
\usepackage{subfigure}
\usepackage{dsfont}
\usepackage{color}
\usepackage{hyperref}

\begin{document}

\title{Pair symmetry conversion in driven multiband superconductors}

\author{Christopher Triola}
\affiliation{Nordic Institute for Theoretical Physics (NORDITA), Stockholm, Sweden}
\affiliation{Center for Quantum Materials (CQM), KTH and Nordita, Stockholm, Sweden}
\author{Alexander V. Balatsky}
\affiliation{Institute for Materials Science (IMS), Los Alamos National Laboratory, Los Alamos New Mexico 87545, USA}
\affiliation{Nordic Institute for Theoretical Physics (NORDITA), Stockholm, Sweden}
\affiliation{Center for Quantum Materials (CQM), KTH and Nordita, Stockholm, Sweden}
\affiliation{ETH Institute for Theoretical Studies (ETH-ITS), ETH Zurich, 8092 Zurich, Switzerland}

\begin{abstract}
It was recently shown that odd-frequency superconducting pair amplitudes can be induced in conventional superconductors subjected to a spatially nonuniform time-dependent drive. It has also been shown that, in the presence of interband scattering, multiband superconductors will possess bulk odd-frequency superconducting pair amplitudes. In this work we build on these previous results to demonstrate that by subjecting a multiband superconductor with interband scattering to a time-dependent drive even-frequency pair amplitudes can be converted to odd-frequency pair amplitudes and vice versa. We will discuss the physical conditions under which these pair symmetry conversions can be achieved and possible experimental signatures of their presence.      
\end{abstract}


\maketitle
\section{Introduction}
Due to the fermionic nature of electrons, the spatial symmetry ($s$-wave, $p$-wave, $d$-wave, etc.) of a superconducting gap is intimately related to the spin state (singlet or triplet)  of the Cooper pairs making up the condensate. In the limit of equal-time pairing this relationship is quite simple, even-parity gaps (like $s$-wave, or $d$-wave) correspond to spin singlet states while odd-parity gaps (like $p$-wave or $f$-wave) correspond to spin triplet states. However, if the electrons are paired at unequal times the superconducting gap could be odd in time or, equivalently, odd in frequency (odd-$\omega$), in which case the condensate could be even in spatial parity and spin triplet or odd in spatial parity and spin singlet. This possibility, originally posited for $^3$He by Berezinskii\cite{Berezinskii1974} and then later for superconductivity\cite{BalatskyPRB1992}, is intriguing both because of the unconventional symmetries which it permits and for the fact that it represents a class of hidden order, due to the vanishing of equal time correlations. 

While some research has been dedicated to the thermodynamic stability of intrinsically odd-$\omega$ phases\cite{heid1995thermodynamic,solenov2009thermodynamical,kusunose2011puzzle,FominovPRB2015}, a great deal of previous research has been devoted to the identification of heterostructures in which odd-$\omega$ pairing could be induced including: ferromagnetic - superconductor heterostructures \cite{BergeretPRL2001,bergeret2005odd,yokoyama2007manifestation,houzet2008ferromagnetic,EschrigNat2008,LinderPRB2008,crepin2015odd}, topological insulator - superconductor systems \cite{YokoyamaPRB2012,Black-SchafferPRB2012,Black-SchafferPRB2013,TriolaPRB2014}, normal metal - superconductor junctions due to broken translation symmetry\cite{tanaka2007theory,TanakaPRB2007,LinderPRL2009,LinderPRB2010_2,TanakaJPSJ2012}, two-dimensional bilayers coupled to conventional $s$-wave superconductors\cite{parhizgar_2014_prb}, and in generic two-dimensional electron gases coupled to superconductor thin films\cite{triola2016prl}. In addition to theoretical studies, there are experimental indications of the realization of odd-$\omega$ pairing at the interface of Nb thin films and epitaxial Ho\cite{di2015signature}. Furthermore, the concept of odd-$\omega$ order parameters can be generalized to charge and spin density waves\cite{pivovarov2001odd,kedem2015odd} and Majorana fermion pairs\cite{huang2015odd}, demonstrating the pervasiveness of the odd-$\omega$ class of ordered states. 

Additionally, it has been shown that superconductors with multiple bands close to the Fermi level, like MgB$_2$\cite{nagamatsu2001superconductivity,bouquet2001specific,brinkman2002multiband,golubov2002specific,iavarone2002two} and iron-based superconductors\cite{hunte2008two,kamihara2008iron,ishida2009extent,cvetkovic2009multiband,kamihara2008iron,stewart2011superconductivity}, will possess odd-$\omega$ pairing in the presence of interband hybridization\cite{black2013odd,komendova2015experimentally,AsanoPRB2015,komendova2017odd}. An advantage of studying odd-$\omega$ pairing in multiband superconductors is that these systems do not have to be engineered to generate odd-$\omega$ pair amplitudes since interband scattering can arise from disorder or it can be intrinsic to the system if the Cooper pairs are composed of electrons corresponding to particular orbitals while the quasiparticles of the system emerge from a linear combination of these orbitals\cite{komendova2015experimentally}, as is the case in Sr$_2$RuO$_4$\cite{komendova2017odd}. Thus, it is expected that bulk odd-$\omega$ pairing should be ubiquitous in multiband superconductors. 

Motivated by the intrinsically dynamical nature of odd-$\omega$ condensates, we recently demonstrated the possibility of inducing odd-$\omega$ superconducting pair amplitudes in a conventional $s$-wave superconductor in the presence of a spatially non-uniform and time-dependent external electric field\cite{triolaprb2016}. The purpose of our current work is both to extend this result to the case of driven multiband superconductors and to examine the nature of pair symmetry conversion in these systems, establishing a relationship between the symmetry of the dynamically generated pair amplitudes and the symmetry of the pair amplitudes in the absence of a drive. Specifically, we consider a superconductor with two bands close to the Fermi level, each possessing a conventional intraband $s$-wave gap, with a finite interband hybridization so that both even-$\omega$ and odd-$\omega$ pair amplitudes are present. Then, using perturbation theory, we show that, in the presence of a time-dependent drive, novel odd-$\omega$ pair amplitudes are generated from even-$\omega$ amplitudes and novel even-$\omega$ amplitudes are generated from odd-$\omega$ amplitudes. We also demonstrate that the conditions for this dynamical pair symmetry conversion coincide with the conditions for the emergence of certain peak structures in the quasiparticle density of states (DOS). 

It should be noted that, while a great deal of work has been dedicated to inducing odd-$\omega$ pairing in systems with only even-$\omega$ pairing, our study examines the inverse effect: inducing even-$\omega$ pairing from previously-existing odd-$\omega$ pairing. This novel effect offers an additional means to modify the pairing states of existing systems. Furthermore, given that even-$\omega$ states are typically associated with sharp spectral features, this effect could point toward new directions for measuring and quantifying odd-$\omega$ superconducting states.    

The remainder of this paper is organized as follows. In Section~\ref{sec:model}, we establish the model we will use to describe a conventional $s$-wave singlet superconductor with two bands close to the Fermi level, and review the conditions under which interband scattering can lead to odd-$\omega$ pairing. In Section~\ref{sec:pert}, we: derive the corrections to the Green's functions to leading order in the drive amplitude; present the conditions for the conversion of even-$\omega$ pair amplitudes to odd-$\omega$ pair amplitudes and vice versa; and discuss possible signatures in the DOS. In Section~\ref{sec:selfgap}, we account for self-consistent corrections to the gap, demonstrating the robustness of the effect. In Section~\ref{sec:con}, we offer concluding remarks.   

\section{Model}
\label{sec:model}
The physical system we wish to consider is a superconductor in which multiple quasiparticle bands are close to the Fermi level, as is the case in MgB$_2$\cite{nagamatsu2001superconductivity,bouquet2001specific,brinkman2002multiband,golubov2002specific,iavarone2002two} and iron-based superconductors\cite{hunte2008two,kamihara2008iron,ishida2009extent,cvetkovic2009multiband,kamihara2008iron,stewart2011superconductivity}. We assume that the superconductor has an $s$-wave spin singlet order parameter, $\Delta_{\alpha\beta}$, with band indices allowing for pairing in both the interband and intraband channels. Additionally, as in previous studies\cite{black2013odd,komendova2015experimentally,AsanoPRB2015,komendova2017odd}, we account for a phenomenological interband scattering which could be caused by disorder or by a mismatch between the orbital structure of the quasiparticle bands and the superconducting order parameter. In this work we will consider both the case of a two-dimensional (2D) thin film superconductor and a three-dimensional (3D) superconductor. For concreteness, unless otherwise specified, all numerical work will be performed assuming two quasiparticle bands and model parameters associated with the two-band superconductor MgB$_2$. Starting from this system, we will examine the affect of an applied time-dependent drive, which, for concreteness, we assume to be an AC electric field, which could be realized through gating (in the case of a 2D superconductor) or using an RF source.            

To describe this system we employ the model Hamiltonian:
\begin{equation}
H=H_{\text{sc}}+H_t+H_{\text{bath}}+H_{\text{mix}}
\label{eq:ham}
\end{equation}
where $H_{\text{sc}}$ describes the undriven multiband superconductor, $H_t$ is the time-dependent drive, $H_{\text{bath}}$ describes a Fermionic bath held at inverse temperature $\beta$ which allows for a phenomenological treatment of dissipation, and $H_{\text{mix}}$ describes the coupling between the superconductor and the bath. 

We will proceed using a two-band superconductor allowing for both interband and intraband pairing:
\begin{equation}
\begin{aligned}
H_{\text{sc}} &= \sum_{\textbf{k},\sigma}\left(\xi_{a,\textbf{k}} \psi^\dagger_{\sigma,a,\textbf{k}} \psi_{\sigma,a,\textbf{k}} + \xi_{b,\textbf{k}} \psi^\dagger_{\sigma,b,\textbf{k}} \psi_{\sigma,b,\textbf{k}} \right) \\
&+ \sum_{\alpha,\beta,\textbf{k}} \Delta_{\alpha\beta} \psi^\dagger_{\uparrow,\alpha,-\textbf{k}}\psi^\dagger_{\downarrow,\beta,\textbf{k}} + \text{h.c.} \\
&+ \sum_{\textbf{k},\sigma}\Gamma \psi^\dagger_{\sigma,a,\textbf{k}} \psi_{\sigma,b,\textbf{k}} + \text{h.c.}
\end{aligned}
\label{eq:H_sc}
\end{equation}
where $\xi_{\alpha,\textbf{k}}=\tfrac{k^2}{2m_\alpha}-\mu_\alpha$ is the quasiparticle dispersion in band $\alpha$ with effective mass $m_\alpha$ measured from the chemical potential $\mu_\alpha$, $\psi^\dagger_{\sigma,\alpha,\textbf{k}}$ ($\psi_{\sigma,\alpha,\textbf{k}}$) creates (annihilates) a quasiparticle with spin $\sigma$ in band $\alpha$ with momentum $\textbf{k}$, $\Delta_{\alpha\beta}\equiv \lambda \int \tfrac{d^dk}{(2\pi)^d}\langle \psi_{\uparrow,\alpha,-\textbf{k}}\psi_{\downarrow,\beta,\textbf{k}} \rangle$ is the superconducting gap, where $d$ is the dimensionality of the system, and we allow for the possibility of interband scattering with amplitude $\Gamma$. 

With these conventions we write the time-dependent drive as:
\begin{equation}
H_t =\sum_{\textbf{k},\sigma,\alpha,\beta} U_{\alpha\beta}(t)\psi^\dagger_{\sigma,\alpha,\textbf{k}} \psi_{\sigma,\beta,\textbf{k}}.  
\label{eq:H_drive}
\end{equation}
The bath and mixing terms take the form:
\begin{equation}
\begin{aligned}
H_{\text{bath}} &=\sum_{n,\sigma,\alpha,\textbf{k}}\left( \epsilon_{n} -\mu_\text{bath}\right) c^\dagger_{n;\sigma\alpha\textbf{k}} c_{n;\sigma\alpha\textbf{k}} \\
H_{\text{mix}} &=\sum_{\textbf{k},n,\sigma,\alpha}\eta_{n} c^\dagger_{n;\sigma\alpha\textbf{k}} \psi_{\sigma,\alpha,\textbf{k}} + \text{h.c.}
\end{aligned}
\label{eq:H_bath}
\end{equation}
where $\epsilon_{n}$ describes the energy levels of the Fermionic bath, $\mu_\text{bath}$ is the chemical potential of the bath, $c^\dagger_{n;\sigma\alpha\textbf{k}}$ ($c_{n;\sigma\alpha\textbf{k}}$) creates (annihilates) a Fermionic mode with degrees of freedom indexed by $n$, $\sigma$, $\alpha$, and $\textbf{k}$, and $\eta_{n}$ specifies the amplitude of the coupling between the superconductor and the bath. 

From this Hamiltonian we can derive a Dyson equation for the Keldysh Green's functions describing this system (see Appendix~\ref{app:eom} for details):
\begin{widetext}
\begin{equation}
\hat{\mathcal{G}}(\textbf{k};t_1,t_2) = \hat{\mathcal{G}}_0(\textbf{k};t_1-t_2) + \int_{-\infty}^\infty dt \hat{\mathcal{G}}_0(\textbf{k};t_1-t) \left( 
\begin{array}{cc}
\hat{U}(t) & 0 \\
0 & -\hat{U}(t)^*
\end{array}\right)\otimes \hat{\rho}_0 \hat{\mathcal{G}}(\textbf{k};t,t_2)
\label{eq:dyson_keldysh}
\end{equation}
\end{widetext}
where $\hat{\rho}_0$ is the 2$\times$2 identity in Keldysh space, and $\hat{\mathcal{G}}_0(\textbf{k};t_1-t_2)$ is the Green's function describing the undriven system written in the Keldysh basis:
\begin{equation}
\hat{\mathcal{G}}_0(\textbf{k};t_1-t_2)=\left(\begin{array}{cc}
\hat{\mathcal{G}}_0^{\text{R}}(\textbf{k};t_1-t_2) & \hat{\mathcal{G}}_0^{\text{K}}(\textbf{k};t_1-t_2) \\
0 & \hat{\mathcal{G}}_0^{\text{A}}(\textbf{k};t_1-t_2)
\end{array} \right)
\end{equation}
where $\hat{\mathcal{G}}_0^{\text{R}}(\textbf{k};t_1-t_2)$, $\hat{\mathcal{G}}_0^{\text{A}}(\textbf{k};t_1-t_2)$, and $\hat{\mathcal{G}}_0^{\text{K}}(\textbf{k};t_1-t_2)$ are the retarded, advanced, and Keldysh Green's functions, respectively. 

After integrating out the bath (see Appendix~\ref{app:bath}), the Fourier transform of $\hat{\mathcal{G}}_0^{\text{R}}(\textbf{k};t_1-t_2)$ is given by:
\begin{widetext}
\begin{equation}
\left(\begin{array}{cccc}
\omega + i\eta - \xi_{a,\textbf{k}} & -\Gamma & -\Delta_{aa} & -\Delta_{ab} \\
-\Gamma & \omega + i\eta - \xi_{b,\textbf{k}} & -\Delta_{ba} & -\Delta_{bb} \\
-\Delta_{aa}^* & -\Delta_{ba}^* & \omega + i\eta + \xi_{a,\textbf{k}} & \Gamma \\
-\Delta_{ab}^* & -\Delta_{bb}^* & \Gamma & \omega + i\eta +\xi_{b,\textbf{k}}
\end{array} \right)\hat{\mathcal{G}}_0^{\text{R}}(\textbf{k};\omega) = \mathds{1}
\label{eq:GR}
\end{equation} 
\end{widetext}
where $\eta$ is a constant related to the DOS of the bath, but, for our purposes, will be treated as a phenomenological parameter describing quasiparticle dissipation. 

In equilibrium, the advanced and Keldysh Green's functions may be obtained from $\hat{\mathcal{G}}_0^{\text{R}}(\textbf{k};\omega)$ by:
\begin{equation}
\begin{aligned}
\hat{\mathcal{G}}^{\text{A}}_0(\textbf{k};\omega)&=\hat{\mathcal{G}}^{\text{R}}_0(\textbf{k};\omega)^\dagger \\
\hat{\mathcal{G}}^{\text{K}}_0(\textbf{k};\omega)&=\tanh\left(\frac{\beta\omega}{2} \right)\left[\hat{\mathcal{G}}^{\text{R}}_0(\textbf{k};\omega) - \hat{\mathcal{G}}^{\text{A}}_0(\textbf{k};\omega) \right]
\end{aligned}
\end{equation}
where $\beta$ is the inverse temperature of the bath.

\subsection{Odd-Frequency Pairing From Interband Scattering}
The emergence of odd-$\omega$ pairing in multiband superconductors due to interband scattering has previously been studied \cite{black2013odd,komendova2015experimentally}. One way to see the emergence of these odd-$\omega$ terms is to consider the simple case in which $\Delta_{ab}=\Delta_{ba}=0$ and solve for $\hat{\mathcal{G}}_0^{\text{R}}(\textbf{k};\omega)$ using Eq (\ref{eq:GR}) which, in the limit of $\eta\rightarrow 0$, is given by:
\begin{equation}
\hat{\mathcal{G}}_0^{\text{R}}(\textbf{k};\omega) = \left(\begin{array}{cc} \hat{G}_0^{\text{R}}(\textbf{k};\omega) & \hat{F}_0^{\text{R}}(\textbf{k};\omega) \\
\hat{\overline{F}}_0^{\text{R}}(\textbf{k};\omega) & \hat{\overline{G}}_0^{\text{R}}(\textbf{k};\omega)
\end{array} \right)
\end{equation} 
where
\begin{widetext}
\begin{equation}
\begin{aligned}
\hat{G}_0^{\text{R}}(\textbf{k};\omega) &=g(\textbf{k},\omega)\left(\begin{array}{cc}
\left(\omega+\xi_{a,\textbf{k}}\right)\left(\omega^2-E_{b,\textbf{k}}^2\right)-\Gamma^2\left(\omega-\xi_{b,\textbf{k}}\right) & \Gamma\left[\left(\omega+\xi_{a,\textbf{k}}\right)\left(\omega+\xi_{b,\textbf{k}}\right)-\Gamma^2-\Delta_{aa}\Delta_{bb}\right] \\
\Gamma\left[\left(\omega+\xi_{a,\textbf{k}}\right)\left(\omega+\xi_{b,\textbf{k}}\right)-\Gamma^2-\Delta_{aa}\Delta_{bb}\right] & \left(\omega+\xi_{b,\textbf{k}}\right)\left(\omega^2-E_{a,\textbf{k}}^2\right)-\Gamma^2\left(\omega-\xi_{a,\textbf{k}}\right)
\end{array} \right), \\
\hat{F}_0^{\text{R}}(\textbf{k};\omega) &=g(\textbf{k},\omega)\left(\begin{array}{cc}
\Delta_{aa}\left(\omega^2-E_{b,\textbf{k}}^2\right)-\Delta_{bb}\Gamma^2 & \Gamma\left[-\omega\left(\Delta_{aa}-\Delta_{bb}\right)+\xi_{a,\textbf{k}}\Delta_{bb}+\xi_{b,\textbf{k}}\Delta_{aa}\right] \\
\Gamma\left[\omega\left(\Delta_{aa}-\Delta_{bb}\right)+\xi_{a,\textbf{k}}\Delta_{bb}+\xi_{b,\textbf{k}}\Delta_{aa}\right] & \Delta_{bb}\left(\omega^2-E_{a,\textbf{k}}^2\right)-\Delta_{aa}\Gamma^2
\end{array}\right)
\end{aligned}
\label{eq:GR_0}
\end{equation}
where we have defined
\begin{equation}
\begin{aligned}
g(\textbf{k},\omega)&=\frac{1}{\left[\omega^2-\epsilon_{+}(\textbf{k})^2 \right]\left[\omega^2-\epsilon_{-}(\textbf{k})^2 \right]}, \\
E_{\alpha,\textbf{k}}&=\sqrt{\xi_{\alpha,\textbf{k}}^2+\Delta_{\alpha\alpha}^2}, \\
\epsilon_{\pm}(\textbf{k})&=\sqrt{\frac{E_{a,\textbf{k}}^2+E_{b,\textbf{k}}^2}{2}+\Gamma^2\pm\sqrt{\left(\frac{E_{a,\textbf{k}}^2-E_{b,\textbf{k}}^2}{2}\right)^2+\Gamma^2\left(\xi_{a,\textbf{k}}+\xi_{b,\textbf{k}}\right)^2+\Gamma^2\left(\Delta_{aa}-\Delta_{bb}\right)^2  } }.
\end{aligned}
\label{eq:spec}
\end{equation}
\end{widetext}
From these expressions one can find $\hat{\overline{G}}^{\text{R}}_0(\textbf{k};\omega)$ and $\hat{\overline{F}}^{\text{R}}_0(\textbf{k};\omega)$ using the definitions:
\begin{equation}
\begin{aligned}
G^{\text{R}}_0(1;2)&=-i\theta(t_1-t_2)\langle \{ \psi_{\alpha_1,\textbf{r}_1}(t_1),\psi^\dagger_{\alpha_2,\textbf{r}_2}(t_2)\}\rangle \\
\overline{G}^{\text{R}}_0(1;2)&=-i\theta(t_1-t_2)\langle \{ \psi^\dagger_{\alpha_1,\textbf{r}_1}(t_1),\psi_{\alpha_2,\textbf{r}_2}(t_2)\}\rangle \\
F^{\text{R}}_0(1;2)&=-i\theta(t_1-t_2)\langle \{ \psi_{\alpha_1,\textbf{r}_1}(t_1),\psi_{\alpha_2,\textbf{r}_2}(t_2)\}\rangle \\
\overline{F}^{\text{R}}_0(1;2)&=-i\theta(t_1-t_2)\langle \{ \psi^\dagger_{\alpha_1,\textbf{r}_1}(t_1),\psi^\dagger_{\alpha_2,\textbf{r}_2}(t_2)\}\rangle
\end{aligned}
\label{eq:bargs}
\end{equation} 
with which one can show $\hat{\overline{G}}^{\text{R}}_0(\textbf{k};\omega)=-\hat{G}_0^{\text{R}}(-\textbf{k};-\omega)^*$ and $\hat{\overline{F}}^{\text{R}}_0(\textbf{k};\omega)=-\hat{F}_0^{\text{R}}(-\textbf{k};-\omega)^*$. Notice, that, because $g(\textbf{k},\omega)=g(\textbf{k},-\omega)$, in Eq (\ref{eq:GR_0}) the interband scattering ($\Gamma\neq 0$) has induced a finite odd-$\omega$ interband pairing in $\hat{F}_0^{\text{R}}(\textbf{k};\omega)$, as shown previously\cite{black2013odd}. 

We will now use these expressions to demonstrate that the presence of a time-dependent drive will not only induce similar odd-$\omega$ terms but also generate additional even-$\omega$ terms as a direct consequence of the odd-$\omega$ terms in Eq (\ref{eq:GR_0}).  

\section{Perturbative Analysis}
\label{sec:pert}
Iterating the Dyson equation in Keldysh space, Eq (\ref{eq:dyson_keldysh}), one can obtain the components of the Green's function to linear order in the drive:
\begin{widetext}
\begin{equation}
\hat{\mathcal{G}}(\textbf{k};t_1,t_2) = \hat{\mathcal{G}}_0(\textbf{k};t_1-t_2) + \int_{-\infty}^\infty dt \hat{\mathcal{G}}_0(\textbf{k};t_1-t) \left( 
\begin{array}{cc}
\hat{U}(t) & 0 \\
0 & -\hat{U}(t)^*
\end{array}\right)\otimes \hat{\rho}_0 \hat{\mathcal{G}}_0(\textbf{k};t-t_2).
\label{eq:linear_t}
\end{equation}
Fourier transforming with respect to the relative ($t_1-t_2$) and average ($(t_1+t_2)/2$) times we can obtain the linear order corrections in frequency space:
\begin{equation}
\hat{\mathcal{G}}(\textbf{k};\omega,\Omega) = 2\pi\delta(\Omega)\hat{\mathcal{G}}_0(\textbf{k};\omega) +\hat{\mathcal{G}}_0(\textbf{k};\omega+\tfrac{\Omega}{2}) \left( 
\begin{array}{cc}
\hat{U}(\Omega) & 0 \\
0 & -\hat{U}(-\Omega)^*
\end{array}\right)\otimes \hat{\rho}_0 \hat{\mathcal{G}}_0(\textbf{k};\omega-\tfrac{\Omega}{2}).
\label{eq:linear_w}
\end{equation}
Focusing on the anomalous part of the Green's functions, we find the terms to linear order in the drive are given by:
\begin{equation}
\begin{aligned}
\delta \hat{F}^{\text{R}}(\textbf{k};\omega,\Omega)&=\hat{G}_0^{\text{R}}(\textbf{k};\omega+\tfrac{\Omega}{2})\hat{U}(\Omega)\hat{F}_0^{\text{R}}(\textbf{k};\omega-\tfrac{\Omega}{2}) -\hat{F}_0^{\text{R}}(\textbf{k};\omega+\tfrac{\Omega}{2})\hat{U}^*(-\Omega)\hat{\overline{G}}_0^{\text{R}}(\textbf{k};\omega-\tfrac{\Omega}{2}) \\
\delta \hat{F}^{\text{A}}(\textbf{k};\omega,\Omega)&=\hat{G}_0^{\text{A}}(\textbf{k};\omega+\tfrac{\Omega}{2})\hat{U}(\Omega)\hat{F}_0^{\text{A}}(\textbf{k};\omega-\tfrac{\Omega}{2}) -\hat{F}_0^{\text{A}}(\textbf{k};\omega+\tfrac{\Omega}{2})\hat{U}^*(-\Omega)\hat{\overline{G}}_0^{\text{A}}(\textbf{k};\omega-\tfrac{\Omega}{2}) \\
\delta \hat{F}^{\text{K}}(\textbf{k};\omega,\Omega)&=\hat{G}_0^{\text{R}}(\textbf{k};\omega+\tfrac{\Omega}{2})\hat{U}(\Omega)\hat{F}_0^{\text{K}}(\textbf{k};\omega-\tfrac{\Omega}{2}) -\hat{F}_0^{\text{R}}(\textbf{k};\omega+\tfrac{\Omega}{2})\hat{U}^*(-\Omega)\hat{\overline{G}}_0^{\text{K}}(\textbf{k};\omega-\tfrac{\Omega}{2}) \\
&+\hat{G}_0^{\text{K}}(\textbf{k};\omega+\tfrac{\Omega}{2})\hat{U}(\Omega)\hat{F}_0^{\text{A}}(\textbf{k};\omega-\tfrac{\Omega}{2}) -\hat{F}_0^{\text{K}}(\textbf{k};\omega+\tfrac{\Omega}{2})\hat{U}^*(-\Omega)\hat{\overline{G}}_0^{\text{A}}(\textbf{k};\omega-\tfrac{\Omega}{2}). 
\end{aligned}
\label{eq:dFRAK}
\end{equation}
\end{widetext}

To demonstrate the emergence of the even-$\omega$ and odd-$\omega$ terms we will focus on the retarded components of the anomalous Green's functions in Eq (\ref{eq:dFRAK}). In general, these corrections, $\delta\hat{F}^{\text{R}}(\textbf{k};\omega,\Omega)$, could possess terms that are even in $\omega$ and terms that are odd in $\omega$. To separate these two possibilities we define:
\begin{equation}
\delta \hat{F}^{\text{e}/\text{o}}(\textbf{k};\omega,\Omega)=\frac{\delta \hat{F}^{\text{R}}(\textbf{k};\omega,\Omega)\pm\delta \hat{F}^{\text{R}}(\textbf{k};-\omega,\Omega)}{2}.
\label{eq:dFR_even_odd}
\end{equation}
By inserting the expressions for the undriven Green's functions into Eq (\ref{eq:dFRAK}) and evaluating Eq (\ref{eq:dFR_even_odd}), one can study the conditions under which new even-$\omega$ pair amplitudes, $\delta \hat{F}^{\text{e}}(\textbf{k};\omega,\Omega)$, and new odd-$\omega$ pair amplitudes, $\delta \hat{F}^{\text{o}}(\textbf{k};\omega,\Omega)$, will be generated in the presence of the drive. The general expressions are quite complicated, therefore we will begin our analysis by studying the simple case in which no odd-$\omega$ amplitudes are present in the undriven system. 

\subsection{Odd-Frequency in Driven Multiband Superconductor for $\Gamma=0$}

In the absence of interband scattering, $\Gamma=0$, the anomalous Green's function of the undriven superconductor, Eq (\ref{eq:GR_0}), possesses only even-$\omega$ terms. To see under what conditions the application of a drive will induce odd-$\omega$ pairing we substitute Eq (\ref{eq:GR_0}) into Eqs (\ref{eq:dFRAK}) and (\ref{eq:dFR_even_odd}) and we find that the odd-$\omega$ corrections to the anomalous Green's function are:
\begin{widetext}
\begin{equation}
\begin{aligned}
\delta F^{\text{o}}_{\alpha\beta}(\textbf{k};\omega,\Omega)&=-\omega U_{\alpha\beta}(\Omega) A_{\alpha\beta}(\textbf{k},\omega,\Omega)\left\{ \left(\Delta_{\alpha}-\Delta_{\beta}\right)\left[\left(\omega^2 +\tfrac{\Omega^2}{4}-E_{\alpha,\textbf{k}}^2\right)\left(\omega^2 +\tfrac{\Omega^2}{4}-E_{\beta,\textbf{k}}^2\right)-\omega^2\Omega^2\right] \right. \\ 
&+\left. \Omega\left(E_{\alpha,\textbf{k}}^2-E_{\beta,\textbf{k}}^2\right)\left(\frac{\Omega}{2}\left(\Delta_\alpha + \Delta_\beta\right) + \xi_{\alpha,\textbf{k}}\Delta_\beta +\xi_{\beta,\textbf{k}}\Delta_\alpha \right) \right\}
\end{aligned}
\label{eq:driven_odd}
\end{equation}  
where 
\begin{equation}
 A_{\alpha\beta}(\textbf{k},\omega,\Omega)=\frac{1}{\left[\left(\omega +\tfrac{\Omega}{2} \right)^2-\xi^2_{\alpha,\textbf{k}}-\Delta^2_\alpha\right]\left[\left(\omega -\tfrac{\Omega}{2} \right)^2-\xi^2_{\alpha,\textbf{k}}-\Delta^2_\alpha\right]\left[\left(\omega +\tfrac{\Omega}{2} \right)^2-\xi^2_{\beta,\textbf{k}}-\Delta^2_\beta\right]\left[\left(\omega -\tfrac{\Omega}{2} \right)^2-\xi^2_{\beta,\textbf{k}}-\Delta^2_\beta\right]}.
\end{equation}
\end{widetext}
From Eq (\ref{eq:driven_odd}) we can see that odd-$\omega$ pairing will emerge in the limit of a static drive, $U_{\alpha\beta}(\Omega)=U_{\alpha\beta}\delta(\Omega)$, only if $\hat{U}$ is off-diagonal in the band index, consistent with previous results for multiband superconductors\cite{black2013odd,komendova2015experimentally,AsanoPRB2015}. However, when $\hat{U}$ is time-dependent an additional term in Eq (\ref{eq:driven_odd}) emerges, proportional to $\Omega$. As with the static case, this term is only nonzero if $\hat{U}(\Omega)$ is off-diagonal in the band index. However, unlike the static case, the dynamical contribution can be nonzero even if the two gaps are equal so long as the two bands have different dispersions.

This result is a simple example of the phenomenon of dynamical pair symmetry conversion, whereby even-$\omega$ pairing amplitudes are converted to odd-$\omega$ amplitudes in the presence of a time-dependent drive. We will now investigate the more general case, in which both even-$\omega$ and odd-$\omega$ pairing amplitudes are already present before the drive is turned on and the application of a time-dependent drive converts the odd-$\omega$ amplitudes to even-$\omega$ amplitudes and vice versa.

\subsection{Symmetry Conversion in Driven Multiband Superconductor for $\Gamma\neq0$}
\label{sec:oddeven}

When interband scattering is allowed, $\Gamma\neq 0$, the anomalous Green's function of the multiband superconductor will possess both odd-$\omega$ terms and even-$\omega$ terms, even in the absence of a time-dependent drive. To distinguish between these ambient odd-$\omega$ and even-$\omega$ components it is useful to define:
\begin{equation}
\hat{F}^{(\text{e}/\text{o})}(\textbf{k};\omega)=\frac{\hat{F}_{0}^{\text{R}}(\textbf{k};\omega) \pm \hat{F}_{0}^{\text{R}}(\textbf{k};-\omega)}{2}.
\label{eq:FR_even_odd}
\end{equation}
By substituting Eq (\ref{eq:FR_even_odd}) into Eqs (\ref{eq:dFRAK}) and (\ref{eq:dFR_even_odd}) we can show that the even-$\omega$ corrections to the anomalous Green's function due to the time-dependent drive are given by:
\begin{equation}
\delta \hat{F}^{\text{e}}(\textbf{k};\omega,\Omega)= \delta F_{\text{e}\rightarrow\text{e}}(\textbf{k};\omega,\Omega) + \delta F_{\text{o}\rightarrow\text{e}} (\textbf{k};\omega,\Omega)
\label{dFR_even}
\end{equation}
and the odd-$\omega$ corrections are given by:
\begin{equation}
\delta \hat{F}^{\text{o}}(\textbf{k};\omega,\Omega)= \delta F_{\text{o}\rightarrow\text{o}}(\textbf{k};\omega,\Omega) + \delta F_{\text{e}\rightarrow\text{o}} (\textbf{k};\omega,\Omega) 
\label{dFR_odd}
\end{equation}
where we have isolated the corrections which preserve frequency parity:
\begin{equation}
\begin{aligned}
\delta F_{\text{e}\rightarrow\text{e}}(\textbf{k};\omega,\Omega)&=\left[ \hat{G}^{\text{R}}_0\left(\textbf{k};\omega+\tfrac{\Omega}{2}\right) \hat{U}(\Omega),\hat{F}^{(\text{e})}\left(\textbf{k};\omega-\tfrac{\Omega}{2}\right) \right]_{+} \\
&+\left[ \hat{G}^{\text{R}}_0\left(\textbf{k};-\omega+\tfrac{\Omega}{2}\right) \hat{U}(\Omega),\hat{F}^{(\text{e})}\left(\textbf{k};\omega+\tfrac{\Omega}{2}\right) \right]_{+}, \\
\delta F_{\text{o}\rightarrow\text{o}}(\textbf{k};\omega,\Omega)&=\left[ \hat{G}^{\text{R}}_0\left(\textbf{k};\omega+\tfrac{\Omega}{2}\right) \hat{U}(\Omega),\hat{F}^{(\text{o})}\left(\textbf{k};\omega-\tfrac{\Omega}{2}\right) \right]_{+} \\
&+\left[ \hat{G}^{\text{R}}_0\left(\textbf{k};-\omega+\tfrac{\Omega}{2}\right) \hat{U}(\Omega),\hat{F}^{(\text{o})}\left(\textbf{k};\omega+\tfrac{\Omega}{2}\right) \right]_{+},
\end{aligned}
\label{eq:parity_preserve}
\end{equation}
and the corrections which reverse frequency parity:
\begin{equation}
\begin{aligned}
\delta F_{\text{e}\rightarrow\text{o}}(\textbf{k};\omega,\Omega)&=\left[ \hat{G}^{\text{R}}_0\left(\textbf{k};\omega+\tfrac{\Omega}{2}\right) \hat{U}(\Omega),\hat{F}^{(\text{e})}\left(\textbf{k};\omega-\tfrac{\Omega}{2}\right) \right]_{-} \\
&-\left[ \hat{G}^{\text{R}}_0\left(\textbf{k};-\omega+\tfrac{\Omega}{2}\right) \hat{U}(\Omega),\hat{F}^{(\text{e})}\left(\textbf{k};\omega+\tfrac{\Omega}{2}\right) \right]_{-}, \\
\delta F_{\text{o}\rightarrow\text{e}}(\textbf{k};\omega,\Omega)&=\left[ \hat{G}^{\text{R}}_0\left(\textbf{k};\omega+\tfrac{\Omega}{2}\right) \hat{U}(\Omega),\hat{F}^{(\text{o})}\left(\textbf{k};\omega-\tfrac{\Omega}{2}\right) \right]_{-} \\
&-\left[ \hat{G}^{\text{R}}_0\left(\textbf{k};-\omega+\tfrac{\Omega}{2}\right) \hat{U}(\Omega),\hat{F}^{(\text{o})}\left(\textbf{k};\omega+\tfrac{\Omega}{2}\right) \right]_{-},
\end{aligned}
\label{eq:parity_reverse}
\end{equation}
where, for convenience, we have defined the bracket:
\begin{equation}
\begin{aligned}
\left[\hat{g}(\omega_1)\hat{u}(\omega_2),\hat{f}(\omega_3) \right]_{\pm}&\equiv \frac{1}{2}\left( \hat{g}(\omega_1)\hat{u}(\omega_2)\hat{f}(\omega_3)\right. \\
&\left. \pm\hat{f}(\omega_3)\hat{u}(-\omega_2)^*\hat{g}(\omega_1)^*\right).
\end{aligned}
\label{eq:bracket}
\end{equation}

From Eqs (\ref{dFR_even})-(\ref{eq:parity_reverse}) we can see that the presence of a time-dependent drive will, in general, generate additional even-$\omega$ and odd-$\omega$ terms in the anomalous Green's function of a multiband superconductor. However, these additional terms could have their origin either from modifying existing correlations with the same symmetry or from symmetry conversion of terms with the opposite frequency parity, i.e. even-$\omega$ terms generating odd-$\omega$ terms or vice versa. To demonstrate that, in general, both symmetry-preserving and symmetry-reversing terms will be nonzero we will now evaluate Eqs (\ref{eq:parity_preserve}) and (\ref{eq:parity_reverse}), explicitly, using Eqs (\ref{eq:GR_0}). 

Assume, for simplicity, that the time-dependent drive takes the form:
\begin{equation}
\hat{U}(\omega)=\left(\begin{array}{cc}
U_0(\omega) & 0 \\
0 & U_0(\omega)
\end{array} \right)
\label{eq:drive}
\end{equation}
where $U_{0}(\omega)$ is given by:
\begin{equation}
\begin{aligned}
U_0(\omega)=2\pi U_0\left[\delta(\omega-\Omega_0) + \delta(\omega+\Omega_0)\right] \\
\end{aligned}
\label{eq:drives}
\end{equation}
which corresponds to a drive proportional to $\cos(\Omega_0t)$ in the time domain. To capture the average time-dependence and relative frequency-dependence we will work with the Wigner transform of the Green's functions, defined as:
\begin{equation}  
\hat{\mathcal{G}}(\textbf{k};\omega,T)=\int \frac{d\Omega}{2\pi}e^{-i\Omega T}\hat{\mathcal{G}}(\textbf{k};\omega,\Omega)
\label{eq:wigner}
\end{equation}
and plot these expressions.

\begin{figure}
 \begin{center}
  \centering
  \includegraphics[width=0.5\textwidth]{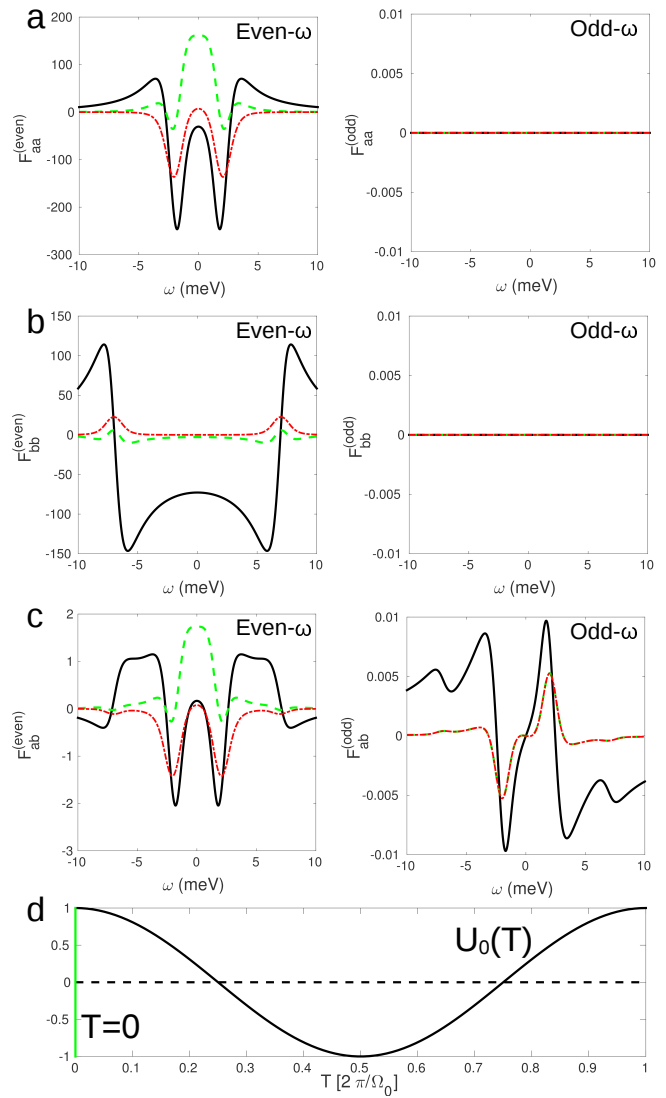}
  \caption{In the left (right) column we plot the even-$\omega$ (odd-$\omega$) terms of the real part of the Wigner transform (defined in Eq (\ref{eq:wigner})) of the anomalous part of the Green's function in Eq (\ref{eq:linear_w}), $\langle \hat{F}^{\text{R}}(\omega,T=0)\rangle$, in black (solid), where we have taken the average value of $\hat{F}^{\text{R}}(\textbf{k};\omega,T=0)$ at $|\textbf{k}|=k^{(a)}_\text{F}$ and $|\textbf{k}|=k^{(b)}_\text{F}$. In each case we have also plotted the parity-preserving terms, Eqs (\ref{eq:parity_preserve}), in green (dashed) and the parity-reversing terms, Eqs (\ref{eq:parity_reverse}), in red (dash-dotted). (a) the diagonal component for band-$a$, (b) the diagonal component for band-$b$, (c) the interband component. (d) The components of the drive from Eq (\ref{eq:drives}) plotted in the time domain over a full period, the green vertical line denotes the time, $T=0$, at which all plots in this figure are evaluated. The parameters used to describe the driven multiband superconductor in this case are: effective masses, $m_a=0.5$ \AA$^{-2}$/eV and $m_b=1$ \AA$^{-2}$/eV; chemical potentials, $\mu_a=\mu_b=2$eV; $s$-wave gaps, $\Delta_{aa}=2$meV, $\Delta_{bb}=7$meV, $\Delta_{ab}=\Delta_{ba}=0$, consistent with MgB$_2$\cite{choi2002origin}; interband scattering, $\Gamma=10$ meV; dissipation described by $\eta=1$meV; and a drive given by Eq (\ref{eq:drives}) with $U_0=10$meV, and $\Omega_0=1$meV (242 GHz). 
         }
  \label{fig:even_odd_u0_10meV_T0}
 \end{center}
\end{figure}

In Fig \ref{fig:even_odd_u0_10meV_T0}, we plot the Wigner transform, at $T=0$, of both the even-$\omega$ and odd-$\omega$ terms of the anomalous Green's function, $\hat{F}^{\text{R}}(\textbf{k};\omega,T)$, for a driven multiband superconductor described by Eqs (\ref{eq:GR_0}) and (\ref{eq:linear_w}) where we have chosen $\Delta_{aa}=2$meV, $\Delta_{bb}=7$meV and $\Gamma=10$meV. We have used an external drive given by Eq (\ref{eq:drives}) with $U_0=10$meV, and $\Omega_0=1$meV. In Fig \ref{fig:even_odd_u0_10meV_T0} we have also included plots of the Wigner transforms of both the symmetry-preserving corrections, Eq (\ref{eq:parity_preserve}), (green/dashed) and the symmetry-reversing corrections, Eq (\ref{eq:parity_reverse}), (red/dash-dotted) to examine the origin of the new contributions. In each plot, in order to show the frequency dependence at the Fermi surface, we have taken the average value of each function evaluated at the two momenta, $|\textbf{k}|=k^{(a)}_{F}=\sqrt{2m_a\mu}$ and $|\textbf{k}|=k^{(b)}_{F}=\sqrt{2m_b\mu}$.

We first turn our attention to the intraband components of the anomalous Green's function, Fig \ref{fig:even_odd_u0_10meV_T0} (a) and (b). Notice that while no new odd-$\omega$ intraband terms are present there are two new contributions to the even-$\omega$ intraband terms, one contribution coming from the ambient even-$\omega$ pairs, and another contribution coming from the ambient odd-$\omega$ pairs. These two contributions are most pronounced in the $F_{aa}$ channel in which they yield a net suppression at $\omega=0$ and a net enhancement at $\omega\approx\pm\Delta_{aa}$. 

Next we consider the interband components of the anomalous Green's function, Fig \ref{fig:even_odd_u0_10meV_T0} (c). Notice a clear enhancement of the odd-$\omega$ terms coming from both the ambient even-$\omega$ and odd-$\omega$ pairs. Additionally, we find an enhancement of the even-$\omega$ interband amplitudes at $\omega\approx\pm\Delta_{aa}$ and $\omega\approx\pm\Delta_{bb}$ coming from the odd-$\omega$ pairs, along with a notable suppression at $\omega=0$ coming from the even-$\omega$ pairs, similar to the case for the even-$\omega$ intraband channels.

\begin{figure}
 \begin{center}
  \centering
  \includegraphics[width=0.5\textwidth]{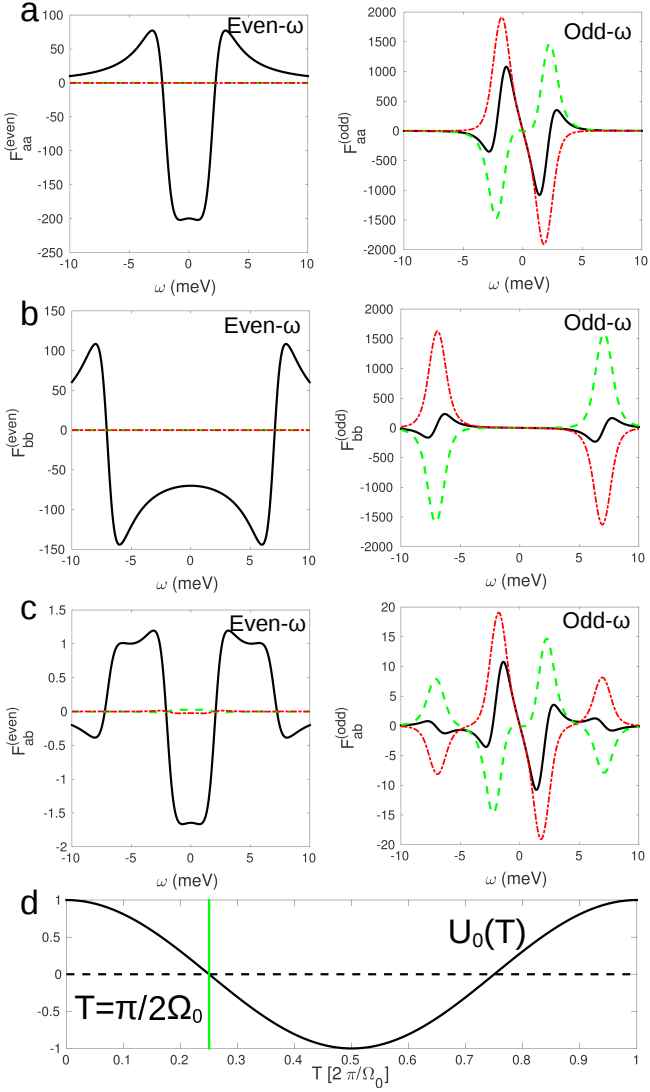}
  \caption{In the left (right) column we plot the even-$\omega$ (odd-$\omega$) terms of the real part of the Wigner transform (defined in Eq (\ref{eq:wigner})) of the anomalous part of the Green's function in Eq (\ref{eq:linear_w}), $\langle \hat{F}^{\text{R}}(\omega,T=\pi/2\Omega_0)\rangle$, in black (solid), where we have taken the average value of $\hat{F}^{\text{R}}(\textbf{k};\omega,T=\pi/2\Omega_0)$ at $|\textbf{k}|=k^{(a)}_\text{F}$ and $|\textbf{k}|=k^{(b)}_\text{F}$. In each case we have also plotted the parity-preserving terms, Eqs (\ref{eq:parity_preserve}), in green (dashed) and the parity-reversing terms, Eqs (\ref{eq:parity_reverse}), in red (dash-dotted). (a) the diagonal component for band-$a$, (b) the diagonal component for band-$b$, (c) the interband component. (d) The components of the drive from Eq (\ref{eq:drives}) plotted in the time domain over a full period, the green vertical line denotes the time, $T=\pi/2\Omega_0$, at which all plots in this figure are evaluated. The parameters used to describe the driven multiband superconductor in this case are: effective masses, $m_a=0.5$ \AA$^{-2}$/eV and $m_b=1$ \AA$^{-2}$/eV; chemical potentials, $\mu_a=\mu_b=2$eV; $s$-wave gaps, $\Delta_{aa}=2$meV, $\Delta_{bb}=7$meV, $\Delta_{ab}=\Delta_{ba}=0$, consistent with MgB$_2$\cite{choi2002origin}; interband scattering, $\Gamma=10$ meV; dissipation described by $\eta=1$meV; and a drive given by Eq (\ref{eq:drives}) with $U_0=10$meV, and $\Omega_0=1$meV (242 GHz).
         }
  \label{fig:even_odd_u0_10meV_T_halfpi}
 \end{center}
\end{figure}

In Fig \ref{fig:even_odd_u0_10meV_T_halfpi}, we plot the Wigner transform, at $T=\pi/2\Omega_0$, of both the even-$\omega$ and odd-$\omega$ terms of the anomalous Green's function, $\hat{F}^{\text{R}}(\textbf{k};\omega,T)$, for a driven multiband superconductor using the same parameters as those appearing in Fig \ref{fig:even_odd_u0_10meV_T0}. In contrast to the results at $T=0$, Fig \ref{fig:even_odd_u0_10meV_T0}, we see that at $T=\pi/2\Omega_0$ the drive has very little affect on the even-$\omega$ terms, but a rather strong affect on the odd-$\omega$ terms. In Fig \ref{fig:even_odd_u0_10meV_T_halfpi} (a) and (b), we see that relatively large intraband odd-$\omega$ amplitudes have emerged at $\omega\approx\pm\Delta_{aa}$ and $\omega\approx\pm\Delta_{bb}$ for the $F_{aa}$ and $F_{bb}$ channels, respectively. By examining the red (dash-dotted) and green (dashed) curves we determine that these novel odd-$\omega$ terms have contributions from both the symmetry-preserving terms and symmetry-reversing terms. However, each contribution can be seen to give rise to distinct peak structures in these channels. Turning our attention to Fig \ref{fig:even_odd_u0_10meV_T_halfpi} (c), the interband anomalous Green's function, we can see similar enhancements of the odd-$\omega$ amplitude at $\omega\approx\pm\Delta_{aa}$ and $\omega\approx\pm\Delta_{bb}$. Just as with the intraband channels, the novel interband terms possess both symmetry-preserving and symmetry-reversing contributions.

To better understand the time-dependence of the pairing amplitudes we have compiled a movie showing the same plots as in Figs~\ref{fig:even_odd_u0_10meV_T0} and~\ref{fig:even_odd_u0_10meV_T_halfpi} over a full period of the drive\cite{sm}. From this movie we observe that, at generic times during the period, contributions to the odd-$\omega$ and even-$\omega$ pair amplitudes are non-zero. Furthermore, we can see that the corrections to the odd-$\omega$ amplitudes are largest exactly when the drive vanishes and smallest exactly when the drive reaches its maximum amplitude. On the other hand the corrections to the even-$\omega$ amplitudes behave in the opposite manner, obtaining their largest contribution exactly when the drive is at its maximum amplitude and smallest contribution when the drive vanishes. 

\subsection{Density of States}
\label{sec:dos}

Now that we have established the possibility of pair symmetry conversion in driven multiband superconductors, we would like to discuss an experimental observable that might indicate that such a conversion has occurred. The time-dependent DOS is one such observable which can be measured using scanning tunneling microscopy (STM)\cite{hofer2003theories,tersoff1983theory}. This quantity can be obtained from the retarded Green's function by:
\begin{equation}
\mathcal{N}_d(\omega,T)= -\frac{1}{\pi}\int \frac{d^dk}{(2\pi)^d} \frac{d\Omega}{2\pi} \text{Im}\left\{ e^{-i\Omega T} \text{Tr} \hat{\mathcal{G}}^{\text{R}}(\textbf{k};\omega,\Omega) \right\}
\label{eq:dos}
\end{equation} 
where $d$ is the dimensionality of the system, and $\hat{\mathcal{G}}^{\text{R}}(\textbf{k};\omega,\Omega)$ can be obtained from Eq (\ref{eq:linear_w}).

\begin{figure}
 \begin{center}
  \centering
  \includegraphics[width=0.5\textwidth]{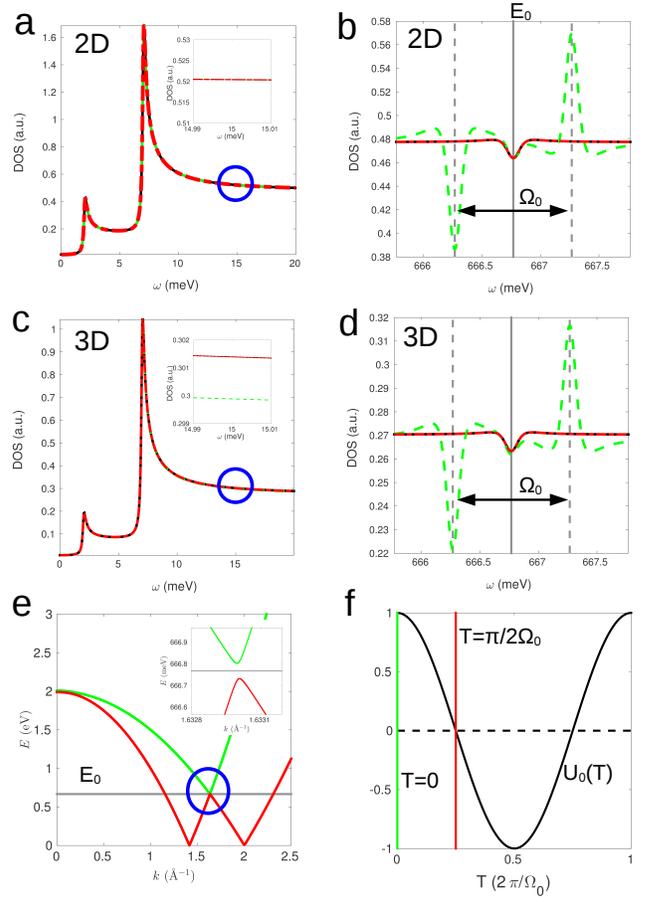}
  \caption{In (a) and (b), the 2D DOS computed using: effective masses, $m_a=0.5$ \AA$^{-2}$/eV and $m_b=1$ \AA$^{-2}$/eV; chemical potentials, $\mu_a=\mu_b=2$eV; $s$-wave gaps, $\Delta_{aa}=2$meV, $\Delta_{bb}=7$meV, $\Delta_{ab}=\Delta_{ba}=0$, consistent with MgB$_2$\cite{choi2002origin}; interband scattering, $\Gamma=10$ meV; dissipation described by $\eta=0.1$meV; and a drive given by Eq (\ref{eq:drives}) with $U_0=10$meV, and $\Omega_0=1$meV (242 GHz). In both panels we show the case for no drive in black (solid), and the cases with the drive at times $T=0$ and $T=\pi/2\Omega_0$ in green (dashed) and red (dash-dotted), respectively. In (a) we focus on the states near the Fermi surface, in (b) we focus on the range of energies near the crossing of the two bands at which we find the driven DOS at $T=0$ possesses two peaks shifted from the avoided crossing at $E_0$ by, $\pm\Omega_0/2$. In (c) and (d), the 3D DOS plotted for the same parameters as in (a) and (b). Notice that the main difference is that in 3D the driven DOS at $T=0$ is slightly suppressed relative to the undriven DOS (see inset). In (e) we plot the spectrum of the two band superconductor given by $\epsilon_{\pm}(\textbf{k})$ in Eq (\ref{eq:spec}). The horizontal grey line denotes the avoided crossing (see inset) at $E_0$, Eq (\ref{eq:peak}), due to the finite interband scattering, $\Gamma$. In (f) we show the drive from Eq (\ref{eq:drives}) plotted in the time domain over a full period, the green vertical line denotes the beginning of the period at $T=0$ where the drive has maximum amplitude, while the red line denotes $T=\pi/2\Omega_0$ where the drive amplitude is zero.
         }
  \label{dos_plots}
 \end{center}
\end{figure}

In Fig~\ref{dos_plots}, we plot $\mathcal{N}_d(\omega,T)$ as a function of frequency, $\omega$, for $d=2$, (a) and (b), and $d=3$, (c) and (d), using the same parameters as in Section~\ref{sec:oddeven}: effective masses, $m_a=0.5 \text{eV}^{-1}\cdot$\AA$^{-2}$ and $m_b=1 \text{eV}^{-1}\cdot$\AA$^{-2}$; chemical potentials, $\mu_a=\mu_b=2$eV; $s$-wave gaps, $\Delta_{aa}=2$meV, $\Delta_{bb}=7$meV, $\Delta_{ab}=\Delta_{ba}=0$; and interband scattering, $\Gamma=10$ meV. However, unlike in Section~\ref{sec:oddeven} we use a dissipation parameter of $\eta=0.1$meV to better highlight the sharp features in the DOS. The black (solid) curves in Figs~\ref{dos_plots}(a)-(d) show $\mathcal{N}_d(\omega,T)=\mathcal{N}_d^{(0)}(\omega)$ without a time-dependent drive, while the green (dashed) and red (dash-dotted) curves show $\mathcal{N}_d(\omega,T)$ in the presence of a drive described by Eq (\ref{eq:drives}) with $U_0=10$meV, and $\Omega_0=1$meV (242 GHz) for times $T=0$ (green/dashed) and $T=\pi/2\Omega_0$ (red/dash-dotted). 

Notice that, for the range of frequencies considered in Fig~\ref{dos_plots}(a) and (c), we see very little difference between the undriven and driven DOS. In each case the dominant features are the coherence peaks associated with the gaps at $\omega\approx |\Delta_{aa}|$ and $|\Delta_{bb}|$ shifted slightly due to the interband scattering, $\Gamma$. In fact, in Fig~\ref{dos_plots}(a) (2D DOS) all curves lie directly on top of each other. However, in Fig~\ref{dos_plots}(c) (3D DOS) the main difference is that for the driven case at $T=0$ there is a slight suppression of the DOS which disappears at $T=\pi/2\Omega_0$ consistent with the fact that the drive in Eq (\ref{eq:drives}) vanishes at $T=\pi/2\Omega_0$. This suppression is a direct consequence of the $\sqrt{\omega}$-dependence of the DOS in 3D (see Appendix~\ref{app:dos}).

In Fig~\ref{dos_plots}(b) and (d), we show the same three DOS curves as in Fig~\ref{dos_plots}(a) and (c) except plotted over a narrow range of frequencies around the avoided crossing in the quasiparticle spectrum of the superconductor (see inset in Fig~\ref{dos_plots}(e)) located at:
\begin{equation}
E_0\approx\sqrt{\Gamma^2 +\mu^2\left(\frac{m_a-m_b}{m_a+m_b}\right)^2+\frac{m_a\Delta_{aa}^2+m_b\Delta_{bb}^2}{m_a+m_b}}.
\label{eq:peak}
\end{equation}  
Notice that for both 2D and 3D the undriven DOS (black curve) exhibits a slight suppression around $E_0$ associated with the depletion of states at the avoided crossing and that the same behavior is exhibited by the driven DOS at $T=\pi/2\Omega_0$. This feature has been noted before in multiband superconductors and shares the same origin as the previously discussed odd-$\omega$ pair amplitudes in multiband superconductors\cite{komendova2015experimentally}, i.e. the interband hybridization. However, at $T=0$ the driven DOS is changed significantly at $E_0$ with two extrema appearing at $E_0\pm\Omega_0/2$, similar to the case in superconductors driven by a spatially nonuniform electric field\cite{triolaprb2016}. The energies associated with these features indicate that their origin can be traced back to the Floquet bands generated by the periodic drive. However, we note that their appearance requires both the presence of a drive and finite interband scattering, necessary and sufficient conditions for the symmetry conversion discussed in Section~\ref{sec:oddeven}. Furthermore, these features can be noticeably enhanced relative to the undriven spectral features at $E_0$, as can be seen from Figs~\ref{dos_plots}(b) and (d). Therefore, we conclude that these peaks offer a potential diagnostic tool for studying pair symmetry conversion in driven multiband superconductors.

\section{Self-Consistent Gap Calculation}
\label{sec:selfgap}
In the previous sections we have demonstrated the possibility of generating both odd- and even-frequency terms in the anomalous Green's function of a multiband superconductor using a time-dependent drive to linear order in the driving amplitude. However, in the above analysis we neglected corrections to the gap function, $\hat{\Delta}$, due to the drive. We will now use the expressions derived in Section \ref{sec:pert} to analyze these additional terms self-consistently and demonstrate the robustness of the effect. For convenience, in this section we will focus on the 3D case.

\begin{figure}[h!]
 \begin{center}
  \centering
  \includegraphics[width=0.5\textwidth]{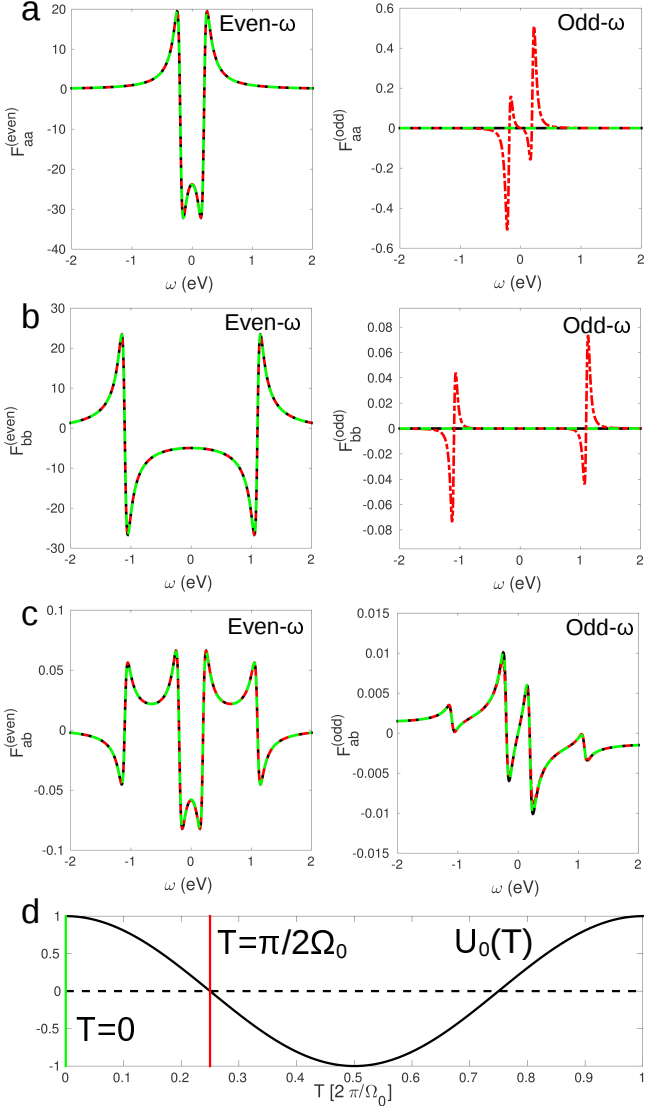}
  \caption{In the left (right) column we plot the even-$\omega$ (odd-$\omega$) terms of the real part of the Wigner transform (defined in Eq (\ref{eq:wigner})) of the anomalous part of the Green's function in Eq (\ref{eq:linear_w}), $\langle \hat{F}^{\text{R}}(\omega,T)\rangle$, where we have taken the average value of $\hat{F}^{\text{R}}(\textbf{k};\omega,T)$ at $|\textbf{k}|=k^{(a)}_\text{F}$ and $|\textbf{k}|=k^{(b)}_\text{F}$. In each panel, $\langle \hat{F}^{\text{R}}(\omega,T)\rangle$ is calculated self-consistently using Eq (\ref{eq:gap}) for the parameters discussed in the text at times $T=0$, green (dashed), and $T=\pi/2\Omega_0$, red (dash-dotted). Additionally, we note that the self-consistent results for these same parameters but with $U_0=0$ appears as a black curve which overlaps almost exactly with the $T=0$ results. (a) the diagonal component for band-$a$, (b) the diagonal component for band-$b$, (c) the interband component. (d) The drive from Eq (\ref{eq:drives}) plotted in the time domain over a full period, the green vertical line denotes the time $T=0$ while the red line denotes $T=\pi/2\Omega_0$.
         }
  \label{fig:even_odd_selfcon_T_halfpi}
 \end{center}
\end{figure}

Assuming the interaction responsible for the superconducting gap is local in relative time and real space, the time-dependent gap is given by:
\begin{equation}
\Delta_{\alpha\beta}(T) = i \lambda \int \frac{d^dk}{(2\pi)^d} \frac{d\omega}{2\pi} \hat{F}^{>}(\textbf{k};\omega,T)
\label{eq:gap}
\end{equation}  
where $\hat{F}^{>}(\textbf{k};\omega,T)$ is the Wigner representation of the $>$ anomalous Green's function which can be expressed in terms of the retarded, advanced, and Keldysh Green's functions:
\begin{equation}
\hat{F}^{>}(\textbf{k};\omega,T)=\frac{1}{2}\left[\hat{F}^{\text{R}}(\textbf{k};\omega,T)-\hat{F}^{\text{A}}(\textbf{k};\omega,T)+\hat{F}^{\text{K}}(\textbf{k};\omega,T) \right].
\label{eq:fgreater}
\end{equation}
This can be expressed in terms of the equilibrium Green's functions, given by Eq (\ref{eq:GR_0}), and the corrections due to the drive:
\begin{equation}
\begin{aligned}
\hat{F}^{>}(\textbf{k};\omega,T)&=\frac{1}{2}\left\{\hat{F}^{\text{R}}_0(\textbf{k};\omega)-\hat{F}^{\text{A}}_0(\textbf{k};\omega)+\hat{F}^{\text{K}}_0(\textbf{k};\omega) \right. \\
&+\left.\delta\hat{F}^{\text{R}}(\textbf{k};\omega,T)-\delta\hat{F}^{\text{A}}(\textbf{k};\omega,T)+\delta\hat{F}^{\text{K}}(\textbf{k};\omega,T) \right\}
\end{aligned}
\label{eq:dfgreater}
\end{equation} 
where these corrections are given by the Wigner transforms of the expressions in Eq (\ref{eq:dFRAK}).

Using Eqs (\ref{eq:gap}) and (\ref{eq:dfgreater}) it is straightforward to compute the components of the gap $\Delta_{\alpha\beta}(T)$ at any average time, $T$, numerically. By inserting these results back into the expressions for $\hat{F}^{>}(\textbf{k};\omega,T)$, recomputing $\Delta_{\alpha\beta}(T)$ and iterating this procedure until the values of $\Delta_{\alpha\beta}(T)$ calculated using Eq (\ref{eq:gap}) match the input values to a precision of our choice we can find self-consistent solutions for the gap in the presence of a drive. 

To illustrate that the effect we have discussed in this paper holds even when the gap is allowed to adjust to the applied time-dependent drive, we have followed the above self-consistent procedure using a precision of $\delta=10^{-5}$ for: effective masses $m_a=1$eV$^{-1}\cdot$\AA$^{-2}$, $m_b=1.5$eV$^{-1}\cdot$\AA$^{-2}$; chemical potentials $\mu_a=\mu_b=10$eV; dissipation parameter $\eta=50$meV; interband scattering $\Gamma=10$meV; intraband drive amplitude $U_0=10$meV; drive frequency $\Omega_0=10$meV (2.4 THz); electron-electron interaction strength $\lambda=1$; and approximately zero temperature. For these parameters the self-consistent gap magnitudes were found at time $T=0$: $|\Delta_{aa}|\approx 197.6$ meV, $|\Delta_{ab}|=|\Delta_{ba}|\approx 1.495$ meV,$|\Delta_{bb}|\approx 1.101$ eV and at time $T= \pi/2\Omega_0$: $|\Delta_{aa}|\approx 197.7$ meV, $|\Delta_{ab}| = |\Delta_{ba}| \approx 1.493$ meV, $|\Delta_{bb} | \approx 1.102$ eV. Additionally, we computed the gaps in the absence of the drive and found precise agreement with the magnitudes at time $T= \pi/2\Omega_0$. Notice that the self-consistent magnitudes do not change appreciably as a function of time $T$; however, there is a slight suppression of the intraband gaps when the drive is at it’s maximum and a slight enhancement of the interband gaps at this same point. Using these self-consistent values for the gaps, we can now examine the frequency-dependent anomalous Green’s functions, and determine whether or not the pair symmetry conversion holds in these cases.

In Fig~\ref{fig:even_odd_selfcon_T_halfpi}(a)-(c) we show the even-$\omega$ and odd-$\omega$ pair amplitudes computed self-consistently for the above parameters plotted as a function of relative frequency, $\omega$, for two different values of the average time, $T$: $T=0$ and $T=\pi/2\Omega_0$. As in Figs~\ref{fig:even_odd_u0_10meV_T0} and \ref{fig:even_odd_u0_10meV_T_halfpi} we have taken the average of $\hat{F}^{\text{R}}(\textbf{k};\omega,T)$ at $|\textbf{k}|=k_{F}^{(a)}$ and $|\textbf{k}|=k_{F}^{(b)}$. First, notice that the intraband odd-$\omega$ terms are only non-negligible at $T=\pi/2\Omega_0$ where they become larger than either of the interband pairing amplitudes. This confirms that the pair symmetry conversion of even-$\omega$ to odd-$\omega$ amplitudes holds even when we account for the corrections to the gap. However, notice that we do not see as dramatic a conversion of odd-$\omega$ to even-$\omega$ amplitudes as we did for the previous cases considered. This is likely because we have restricted ourselves to fairly large equal-time gaps in order to ensure self-consistency in the presence of both interband scattering and a time-dependent drive. 

\begin{figure}[h!]
 \begin{center}
  \centering
  \includegraphics[width=0.5\textwidth]{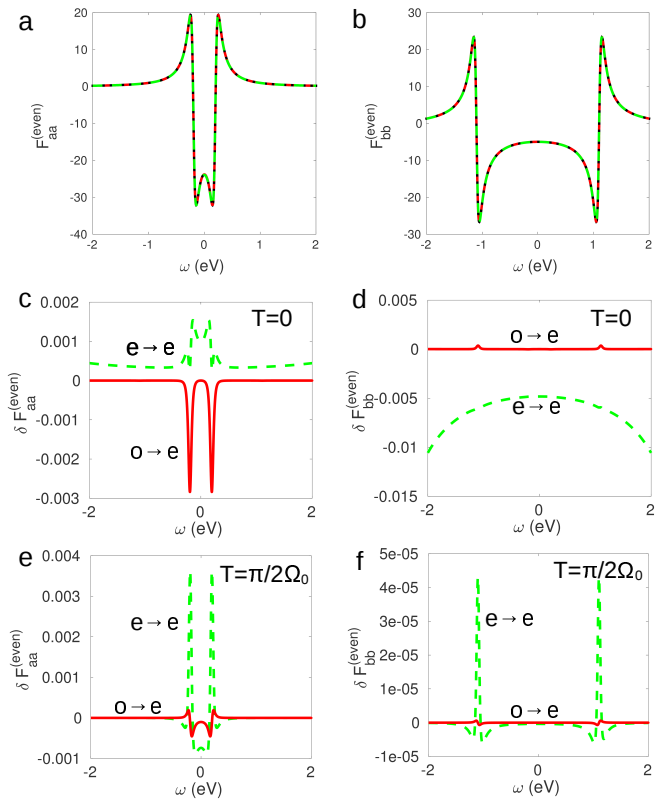}
  \caption{In (a) and (b) we repeat the plots of the even-$\omega$ intraband pair amplitudes appearing in Fig. \ref{fig:even_odd_selfcon_T_halfpi}(a) and \ref{fig:even_odd_selfcon_T_halfpi}(b), which were calculated self-consistently using Eq (\ref{eq:gap}) for the parameters discussed in the text at times $T=0$, green (dashed), $T=\pi/2\Omega_0$, red (dash-dotted), and without a drive black (solid). Plotted over this range, the three curves are essentially indistinguishable; however, there a slight differences which we highlight in (c)-(f). In (c)-(f) we show the symmetry preserving (green/dashed) and  symmetry reversing (red/solid) contributions to the plots appearing in (a) and (b) calculated using Eqs (\ref{eq:parity_preserve}) and (\ref{eq:parity_reverse}). In (c) and (e) we show the corrections to the even-$\omega$ intraband pairing in band-a at times $T=0$ and $T=\pi/2\Omega_0$, respectively; in (d) and (f) we show the corrections to the even-$\omega$ intraband pairing in band-b at times $T=0$ and $T=\pi/2\Omega_0$, respectively.
         }
  \label{fig:even_selfcon}
 \end{center}
\end{figure}

To better understand how the drive affects the even-$\omega$ pair amplitudes, in Figs \ref{fig:even_selfcon}(c)-\ref{fig:even_selfcon}(f) we show both the symmetry preserving (green/dashed) and  symmetry reversing (red/solid) corrections to the even-$\omega$ intraband pair amplitudes appearing in Figs \ref{fig:even_odd_selfcon_T_halfpi}(a) and \ref{fig:even_odd_selfcon_T_halfpi}(b), calculated using Eqs (\ref{eq:parity_preserve}) and (\ref{eq:parity_reverse}). Consistent with the results in Sec III B, we find that, in general, both contributions are nonzero. This confirms that the pair symmetry conversion of odd-$\omega$ to even-$\omega$ amplitudes holds when the self-consistent corrections to the gap are accounted for. In Figs \ref{fig:even_selfcon}(c) and \ref{fig:even_selfcon}(d) we show the even-$\omega$ corrections to the intraband pairing in band-$a$ and band-$b$, respectively, plotted at time $T=0$. Notice, as we found earlier, that the contributions coming from pair symmetry conversion (odd$\rightarrow$even) are strongest at $\omega=|\Delta_{aa}|$ for band-$a$ and $\omega=|\Delta_{bb}|$ for band-$b$. In Figs \ref{fig:even_selfcon}(e) and \ref{fig:even_selfcon}(f) we show the same quantities as Figs \ref{fig:even_selfcon}(c) and \ref{fig:even_selfcon}(d) plotted at time $T=\pi/2\Omega_0$. As we expect from earlier, we see that, at this time, the symmetry reversing contributions are significantly weakened.   

\section{Conclusions}
\label{sec:con}

In this work we considered a model for a two-band superconductor with interband scattering subjected to a time-dependent drive. Working perturbatively, we demonstrated that, not only can the presence of a time-dependent drive be used to generate odd-frequency superconducting pair amplitudes, but also that odd-frequency amplitudes generated from the interband scattering can influence the appearance of the even-frequency amplitudes in the presence of a drive. We have presented a systematic study of the conversion of odd-frequency pair amplitudes to even-frequency pair amplitudes. We also showed that the appearance of the dynamically-induced odd-frequency and even-frequency amplitudes holds even when the gaps are computed self-consistently. Furthermore, by examining the DOS, we found that the conditions for this dynamical pair symmetry conversion also gave rise to novel peak structures, offering a potential signature of the phenomenon.

Since the derivation of the parity-reversing terms, Eq (\ref{eq:parity_reverse}), did not rely on a specific Hamiltonian or gap-symmetry we conclude that these relations should hold in general. These general relations represent a novel means to control the symmetry of Cooper pairs, which could allow for the realization of exotic new superconducting states. Additionally, in light of these results, it would be interesting to study whether or not a time-dependent external field can be used to generate an equal-time gap in an intrinsically odd-frequency superconductor. Since a key feature of odd-frequency superconductors is the vanishing of an equal-time gap, it is conceivable that one could use the kind of pair symmetry conversion proposed in this work to generate sharp spectral features which could expose an otherwise hidden order.

Acknowledgements: We wish to thank David Abergel, Annica Black-Schaffer, Jorge Cayao, Matthias Geilhufe, Yaron Kedem, Lucia Komendov\'a, Sergey Pershoguba, Anna Pertsova, and Enrico Rossi for useful discussions. This work was supported by US DOE BES E3B7, the European Research Council (ERC) DM-321031, and Dr. Max R\"ossler, the Walter Haefner Foundation and the ETH Zurich Foundation.     

\appendix

\section{Derivation of Equations of Motion}
\label{app:eom}
In this appendix, we outline the derivation of the equations of motion describing the Green's functions in Eq (\ref{eq:dyson_keldysh}). 

By commuting the quasiparticle annihilation and creation operators, $\psi_{\sigma,\alpha,\textbf{k}}$ and $\psi^\dagger_{\sigma,\alpha,\textbf{k}}$, with the total Hamiltonian in Eq (\ref{eq:ham}) it is straightforward to derive the Heisenberg equations of motion for these operators:
\begin{equation}
\begin{aligned}
i\frac{d}{dt}\psi_{\sigma,\alpha,\textbf{k}}(t)&=\xi_{\alpha,\textbf{k}}\psi_{\sigma,\alpha,\textbf{k}}(t) \\
&+ \sum_{\alpha'}\left[ \Gamma\left( \hat{\tau}_1\right)_{\alpha\alpha'}+U_{\alpha\alpha'}(t)\right] \psi_{\sigma,\alpha',\textbf{k}}(t) \\
&+\delta_{\sigma\uparrow}\sum_{\alpha'} \Delta_{\alpha\alpha'} \psi^\dagger_{\downarrow,\alpha',-\textbf{k}}(t) \\
&-\delta_{\sigma\downarrow}\sum_{\alpha'} \Delta_{\alpha'\alpha} \psi^\dagger_{\uparrow,\alpha',-\textbf{k}}(t) \\
&+\int_\mathcal{C} dt' \Sigma_{\mathcal{C}}(t-t')\psi_{\sigma,\alpha,\textbf{k}}(t') 
\end{aligned}
\label{a:eq:heom_a}
\end{equation}
and
\begin{equation}
\begin{aligned}
i\frac{d}{dt}\psi^\dagger_{\sigma,\alpha,-\textbf{k}}(t)&=-\xi_{\alpha,-\textbf{k}}\psi^\dagger_{\sigma,\alpha,-\textbf{k}}(t) \\
&- \sum_{\alpha'}\left[ \Gamma\left( \hat{\tau}_1\right)_{\alpha'\alpha}+U_{\alpha\alpha'}^*(t)\right] \psi^\dagger_{\sigma,\alpha',-\textbf{k}}(t) \\
&+\delta_{\sigma\downarrow}\sum_{\alpha'} \Delta^\dagger_{\alpha\alpha'} \psi_{\uparrow,\alpha',\textbf{k}}(t) \\
&-\delta_{\sigma\uparrow}\sum_{\alpha'} \Delta^\dagger_{\alpha'\alpha} \psi_{\downarrow,\alpha',\textbf{k}}(t) \\
&+\int_\mathcal{C} dt' \overline{\Sigma}_{\mathcal{C}}(t-t')\psi^\dagger_{\sigma,\alpha,\textbf{k}}(t') 
\end{aligned}
\label{a:eq:heom_c}
\end{equation}
where $\hat{\tau}_i$ are the Pauli matrices in band space, the contour $\mathcal{C}$ is the standard time contour from the Kadanoff-Baym formalism\cite{rammer2007quantum,maciejko2007introduction,stefanucci2013nonequilibrium,aoki2014nonequilibrium} and we have defined the self-energies associated with the presence of the fermionic bath:
\begin{equation}
\begin{aligned}
\Sigma_{\mathcal{C}}(t-t')&=\sum_{n}\eta_n^2G^{\mathcal{C}}_{\text{bath}}(n;t-t') \\
\overline{\Sigma}_{\mathcal{C}}(t-t')&=\sum_{n}\eta_n^2\overline{G}^{\mathcal{C}}_{\text{bath}}(n;t-t')
\end{aligned}
\end{equation}
where $G^{\mathcal{C}}_{\text{bath}}(n;t-t')=-i\langle \mathcal{T}_\mathcal{C} c_{n;\sigma,\alpha,\textbf{k}}(t)c^\dagger_{n;\sigma,\alpha,\textbf{k}}(t') \rangle$ and $\overline{G}^{\mathcal{C}}_{\text{bath}}(n;t-t')=-i\langle \mathcal{T}_\mathcal{C} c^\dagger_{n;\sigma,\alpha,-\textbf{k}}(t)c_{n;\sigma,\alpha,-\textbf{k}}(t') \rangle$ are contour-ordered Green's functions for the free-fermion bath.

We then define the following contour-ordered Green's functions:
\begin{equation}
\begin{aligned}
G^{\mathcal{C}}_{\sigma_1\alpha_1;\sigma_2\alpha_2}(\textbf{k};t_1,t_2)&=-i\langle \mathcal{T}_\mathcal{C} \psi_{\sigma_1,\alpha_1,\textbf{k}}(t_1)\psi^\dagger_{\sigma_2,\alpha_2,\textbf{k}}(t_2) \rangle \\
\overline{G}^{\mathcal{C}}_{\sigma_1\alpha_1;\sigma_2\alpha_2}(\textbf{k};t_1,t_2)&=-i\langle \mathcal{T}_\mathcal{C} \psi^\dagger_{\sigma_1,\alpha_1,-\textbf{k}}(t_1)\psi_{\sigma_2,\alpha_2,-\textbf{k}}(t_2) \rangle \\
F^{\mathcal{C}}_{\sigma_1\alpha_1;\sigma_2\alpha_2}(\textbf{k};t_1,t_2)&=-i\langle \mathcal{T}_\mathcal{C} \psi_{\sigma_1,\alpha_1,\textbf{k}}(t_1)\psi_{\sigma_2,\alpha_2,-\textbf{k}}(t_2) \rangle \\
\overline{F}^{\mathcal{C}}_{\sigma_1\alpha_1;\sigma_2\alpha_2}(\textbf{k};t_1,t_2)&=-i\langle \mathcal{T}_\mathcal{C} \psi^\dagger_{\sigma_1,\alpha_1,-\textbf{k}}(t_1)\psi^\dagger_{\sigma_2,\alpha_2,\textbf{k}}(t_2) \rangle.
\end{aligned}
\end{equation}
Since the Hamiltonian possesses only trivial spin-dependence we may restrict our attention to the components:
\begin{equation}
\begin{aligned}
G^{\mathcal{C}}_{\alpha_1\alpha_2}(\textbf{k};t_1,t_2)&\equiv G^{\mathcal{C}}_{\uparrow\alpha_1;\uparrow\alpha_2}(\textbf{k};t_1,t_2) \\
\overline{G}^{\mathcal{C}}_{\alpha_1\alpha_2}(\textbf{k};t_1,t_2)&\equiv \overline{G}^{\mathcal{C}}_{\downarrow\alpha_1;\downarrow\alpha_2}(\textbf{k};t_1,t_2) \\
F^{\mathcal{C}}_{\alpha_1\alpha_2}(\textbf{k};t_1,t_2)&\equiv F^{\mathcal{C}}_{\uparrow\alpha_1;\downarrow\alpha_2}(\textbf{k};t_1,t_2) \\
\overline{F}^{\mathcal{C}}_{\alpha_1\alpha_2}(\textbf{k};t_1,t_2)&\equiv \overline{F}^{\mathcal{C}}_{\downarrow\alpha_1;\uparrow\alpha_2}(\textbf{k};t_1,t_2).
\end{aligned}
\end{equation}
Then, using Eqs (\ref{a:eq:heom_a}) and (\ref{a:eq:heom_c}), one can show that these components satisfy the following equations of motion:
\begin{widetext}
\begin{equation}
\left(
\begin{array}{cc}
i\hat{\tau}_0\frac{d}{dt_1} - \hat{h}_{\textbf{k}} -\hat{U}(t_1) & -\hat{\Delta} \\
-\hat{\Delta}^\dagger & i\hat{\tau}_0\frac{d}{dt_1} + \hat{h}_{-\textbf{k}}^* +\hat{U}^*(t_1)
\end{array} \right) \hat{\mathcal{G}}^\mathcal{C}(\textbf{k};t_1,t_2) - \int_\mathcal{C} dt \left(
\begin{array}{cc}
\hat{\tau}_0 \Sigma_\mathcal{C}(t_1-t) & 0 \\
0 & \hat{\tau}_0 \overline{\Sigma}_\mathcal{C}(t_1-t)
\end{array} \right) \hat{\mathcal{G}}^\mathcal{C}(\textbf{k};t,t_2) = \delta_\mathcal{C}(t_1,t_2)
\label{a:eq:contoureom}
\end{equation}
\end{widetext}
where $\hat{\tau}_0$ is the identity matrix in band space, $\hat{h}_{\textbf{k}}$, $\hat{\Delta}$, and $\hat{U}(t)$ are matrices in band-space given by:
\begin{equation}
\begin{aligned}
\hat{h}_{\textbf{k}}&=\left(\begin{array}{cc}
\xi_{a,\textbf{k}} & \Gamma \\
\Gamma & \xi_{b,\textbf{k}} \end{array} \right) \\
\hat{\Delta}&= \left(\begin{array}{cc}
\Delta_{aa} & \Delta_{ab} \\
\Delta_{ba} & \Delta_{bb} \end{array} \right) \\
\hat{U}(t) &= \left(\begin{array}{cc}
U_{aa}(t) & U_{ab}(t) \\
U_{ba}(t) & U_{bb}(t) \end{array} \right)
\end{aligned}
\label{a:eq:bandmatrices}
\end{equation}
and where we define $\hat{\mathcal{G}}^{\mathcal{C}}(\textbf{k};t_1,t_2)$ as:
\begin{equation}
\hat{\mathcal{G}}^\mathcal{C}(\textbf{k};t_1,t_2) = \left(\begin{array}{cc} 
\hat{G}^{\mathcal{C}}(\textbf{k};t_1,t_2) & \hat{F}^{\mathcal{C}}(\textbf{k};t_1,t_2) \\ 
\hat{\overline{F}}^{\mathcal{C}}(\textbf{k};t_1,t_2) & \hat{\overline{G}}^{\mathcal{C}}(\textbf{k};t_1,t_2) \end{array} \right)
\label{a:eq:gcontour}
\end{equation}
where each component is a 2$\times$2 matrix in band space.

Using Eq (\ref{a:eq:contoureom}) it is straightforward to write the Dyson equation for $\hat{\mathcal{G}}^{\mathcal{C}}(\textbf{k};t_1,t_2)$:
\begin{equation}
\begin{aligned}
\hat{\mathcal{G}}^\mathcal{C}(\textbf{k};t_1,t_2) &= \hat{\mathcal{G}}^\mathcal{C}_0(\textbf{k};t_1,t_2) \\
&+ \int_\mathcal{C} dt \hat{\mathcal{G}}^\mathcal{C}_0(\textbf{k};t_1,t) \left( 
\begin{array}{cc}
\hat{U}(t) & 0 \\
0 & -\hat{U}(t)^*
\end{array}\right)\hat{\mathcal{G}}^\mathcal{C}(\textbf{k};t,t_2)
\end{aligned}
\label{a:eq:dyson_contour}
\end{equation}
where $\hat{\mathcal{G}}^\mathcal{C}_0(\textbf{k};t_1,t_2)$ is the solution to Eq (\ref{a:eq:contoureom}) in the absence of a drive.

Assuming that coupling to the bath washes out the correlations between the real and imaginary time contours we may work only with the retarded, advanced, and Keldysh components. Under this assumption we transform Eq (\ref{a:eq:dyson_contour}) to Keldysh space:
\begin{equation}
\begin{aligned}
\hat{\mathcal{G}}(\textbf{k};t_1,t_2) &= \hat{\mathcal{G}}_0(\textbf{k};t_1,t_2) \\
&+ \int_{-\infty}^\infty dt \hat{\mathcal{G}}_0(\textbf{k};t_1,t) \left( 
\begin{array}{cc}
\hat{U}(t) & 0 \\
0 & -\hat{U}(t)^*
\end{array}\right)\otimes \hat{\rho}_0 \hat{\mathcal{G}}(\textbf{k};t,t_2)
\end{aligned}
\label{a:eq:dyson_keldysh}
\end{equation}
where $\hat{\rho}_0$ is the identity in Keldysh space and 
\begin{equation}
\hat{\mathcal{G}}(\textbf{k};t_1,t_2)=\left( \begin{array}{cc}
\hat{\mathcal{G}}^{\text{R}}(\textbf{k};t_1,t_2) & \hat{\mathcal{G}}^{\text{K}}(\textbf{k};t_1,t_2) \\
0 & \hat{\mathcal{G}}^{\text{A}}(\textbf{k};t_1,t_2)
\end{array} \right)
\end{equation}
where each component may be written as linear combinations of the contour-ordered Green's functions:
\begin{widetext}
\begin{equation}
\begin{aligned}
\hat{\mathcal{G}}^{\text{R}}(\textbf{k};t_1,t_2) &= \frac{1}{2}\left[\hat{\mathcal{G}}^{\mathcal{C}}_{11}(\textbf{k};t_1,t_2)-\hat{\mathcal{G}}^{\mathcal{C}}_{12}(\textbf{k};t_1,t_2)+\hat{\mathcal{G}}^{\mathcal{C}}_{21}(\textbf{k};t_1,t_2)-\hat{\mathcal{G}}^{\mathcal{C}}_{22}(\textbf{k};t_1,t_2) \right] \\
\hat{\mathcal{G}}^{\text{A}}(\textbf{k};t_1,t_2) &= \frac{1}{2}\left[\hat{\mathcal{G}}^{\mathcal{C}}_{11}(\textbf{k};t_1,t_2)+\hat{\mathcal{G}}^{\mathcal{C}}_{12}(\textbf{k};t_1,t_2)-\hat{\mathcal{G}}^{\mathcal{C}}_{21}(\textbf{k};t_1,t_2)-\hat{\mathcal{G}}^{\mathcal{C}}_{22}(\textbf{k};t_1,t_2) \right] \\
\hat{\mathcal{G}}^{\text{K}}(\textbf{k};t_1,t_2) &= \frac{1}{2}\left[\hat{\mathcal{G}}^{\mathcal{C}}_{11}(\textbf{k};t_1,t_2)+\hat{\mathcal{G}}^{\mathcal{C}}_{12}(\textbf{k};t_1,t_2)+\hat{\mathcal{G}}^{\mathcal{C}}_{21}(\textbf{k};t_1,t_2)+\hat{\mathcal{G}}^{\mathcal{C}}_{22}(\textbf{k};t_1,t_2) \right]
\end{aligned}
\end{equation} 
\end{widetext}
where $\hat{\mathcal{G}}^{\mathcal{C}}_{ij}(\textbf{k};t_1,t_2)$ is given by the definition in Eq (\ref{a:eq:gcontour}) with the index $i$ ($j$) determining on which path of the contour the time argument $t_1$ ($t_2$) lies, forward $=1$ and backward $=2$ respectively.      

\section{Integrating-Out the Bath}
\label{app:bath} 
In this appendix we outline the procedure for integrating-out the bath and obtaining an expression for the Green's functions in Eq (\ref{eq:GR}). 

In the absence of the drive ($U_{\alpha\beta}(t)=0$) we can Fourier transform Eq (\ref{a:eq:contoureom}) to frequency space to find:
\begin{widetext}
\begin{equation}
\hat{\mathcal{G}}_0(\textbf{k};\omega)  = \left(\begin{array}{cccc}
\hat{\tau}_0\left(\omega-\Sigma^{\text{R}}(\omega)\right)-\hat{h}_\textbf{k} & -\hat{\Delta} & -\hat{\tau}_0 \Sigma^{\text{K}}(\omega) & 0 \\
-\hat{\Delta}^\dagger & \hat{\tau}_0\left(\omega-\overline{\Sigma}^{\text{R}}(\omega)\right)+\hat{h}_{-\textbf{k}} & 0 & -\hat{\tau}_0 \overline{\Sigma}^{\text{K}}(\omega) \\
0 & 0 & \hat{\tau}_0\left(\omega-\Sigma^{\text{A}}(\omega)\right)-\hat{h}_\textbf{k} & -\hat{\Delta} \\
0 & 0 & -\hat{\Delta}^\dagger & \hat{\tau}_0\left(\omega-\overline{\Sigma}^{\text{A}}(\omega)\right)+\hat{h}_{-\textbf{k}}
\end{array}\right) ^{-1}
\label{a:eq:eom_diss}
\end{equation}
\end{widetext}
where 
\begin{equation}
\begin{aligned}
\Sigma^{\text{R}}(\omega)&=\sum_n \eta_n^2 \mathcal{P}\left(\frac{1}{\omega -(\epsilon_n-\mu_{\text{bath}})}\right) \\
&- i\pi\sum_n \eta_n^2  \delta\left(\omega -(\epsilon_n-\mu_{\text{bath}})\right) \\
\overline{\Sigma}^{\text{R}}(\omega)&=\sum_n \eta_n^2 \mathcal{P}\left(\frac{1}{\omega +(\epsilon_n-\mu_{\text{bath}})}\right) \\
&- i\pi\sum_n \eta_n^2  \delta\left(\omega +(\epsilon_n-\mu_{\text{bath}})\right) 
\end{aligned}
\end{equation}
\begin{equation}
\begin{aligned}
\Sigma^{\text{A}}(\omega)&=\sum_n \eta_n^2 \mathcal{P}\left(\frac{1}{\omega -(\epsilon_n-\mu_{\text{bath}})}\right) \\
&+ i\pi\sum_n \eta_n^2  \delta\left(\omega -(\epsilon_n-\mu_{\text{bath}})\right) \\
\overline{\Sigma}^{\text{A}}(\omega)&=\sum_n \eta_n^2 \mathcal{P}\left(\frac{1}{\omega +(\epsilon_n-\mu_{\text{bath}})}\right) \\
&+ i\pi\sum_n \eta_n^2  \delta\left(\omega +(\epsilon_n-\mu_{\text{bath}})\right) 
\end{aligned}
\end{equation}
and
\begin{equation}
\begin{aligned}
\Sigma^{\text{K}}(\omega)&=-i2\pi\tanh\left( \frac{\beta\omega}{2}\right)\sum_n \eta_n^2  \delta\left(\omega -(\epsilon_n-\mu_{\text{bath}})\right) \\
\overline{\Sigma}^{\text{K}}(\omega)&=-i2\pi\tanh\left( \frac{\beta\omega}{2}\right)\sum_n \eta_n^2  \delta\left(\omega +(\epsilon_n-\mu_{\text{bath}})\right).
\end{aligned}
\end{equation}
Assuming a featureless bath we approximate $\eta\approx\pi\sum_n \eta_n^2  \delta\left(\omega -(\epsilon_n-\mu_{\text{bath}})\right)$ and $m\approx\sum_n \eta_n^2 \mathcal{P}\left(\frac{1}{\omega -(\epsilon_n-\mu_{\text{bath}})}\right)$ in which case Eq (\ref{a:eq:eom_diss}) simplifies to:
\begin{widetext}
\begin{equation}
\hat{\mathcal{G}}_0(\textbf{k};\omega)=\left(\begin{array}{cccc}
\hat{\tau}_0\left(\omega+i\eta-m\right)-\hat{h}_\textbf{k} & -\hat{\Delta} & \hat{\tau}_0 i2\tanh\left( \frac{\beta\omega}{2}\right)\eta & 0 \\
-\hat{\Delta}^\dagger & \hat{\tau}_0\left(\omega+i\eta+m\right)+\hat{h}_{-\textbf{k}} & 0 & \hat{\tau}_0 i2\tanh\left( \frac{\beta\omega}{2}\right)\eta \\
0 & 0 & \hat{\tau}_0\left(\omega-i\eta-m\right)-\hat{h}_\textbf{k} & -\hat{\Delta} \\
0 & 0 & -\hat{\Delta}^\dagger & \hat{\tau}_0\left(\omega-i\eta+m\right)+\hat{h}_{-\textbf{k}}
\end{array}\right)^{-1}
\label{a:eq:diss_g0}
\end{equation}
\end{widetext} 
and, without loss of generality, we account for $m$ by shifting the overall chemical potential appearing in $\hat{h}_\textbf{k}$.

\section{Driven Density of States in $d$-dimensions}
\label{app:dos}
In order to illustrate the dependence of the driven DOS on dimension, $d$, we consider a simple model Hamiltonian describing quasiparticles in one-band driven by a time-dependent electric field:
\begin{equation}
H=\sum_{\textbf{k}}\left[ E_\textbf{k} + U(t)\right] \psi^\dagger_{\textbf{k}}\psi_{\textbf{k}}
\end{equation}
where $E_\textbf{k}$ describes the dispersion of the quasiparticles, $U(t)$ is a time-dependent external field, and the momentum is summed over a $d$-dimensional reciprocal space.

Following the exact same reasoning leading to Eq (\ref{eq:linear_w}) one can verify that, to linear order in the drive, the retarded Green's function describing this system is given by:
\begin{equation}
\begin{aligned}
G^{\text{R}}(\textbf{k};\omega,\Omega)&=2\pi\delta(\Omega)G^{\text{R}}_0(\textbf{k};\omega) \\
&+G^{\text{R}}_0(\textbf{k};\omega+\tfrac{\Omega}{2})U(\Omega)G^{\text{R}}_0(\textbf{k};\omega-\tfrac{\Omega}{2})
\end{aligned}
\label{aeq:gr_linear}
\end{equation}
where
\begin{equation}
G^{\text{R}}_0(\textbf{k};\omega)=\lim_{\eta\rightarrow 0}\frac{1}{\omega-E_\textbf{k}+i\eta}.
\label{aeq:gr_0}
\end{equation}
Assuming a drive of the form:
\begin{equation}
U(\Omega)=2\pi U_0\left[\delta(\Omega-\Omega_0)+\delta(\Omega+\Omega_0)\right]
\end{equation}
we may write the Wigner representation of $G^{\text{R}}(\textbf{k};\omega,\Omega)$, which we defined in Eq (\ref{eq:wigner}), as:
\begin{equation}
\begin{aligned}
G^{\text{R}}(\textbf{k};\omega,T)&=G^{\text{R}}_0(\textbf{k};\omega) \\
&+2U_0\cos\left(\Omega_0 T\right)G^{\text{R}}_0(\textbf{k};\omega+\tfrac{\Omega_0}{2})G^{\text{R}}_0(\textbf{k};\omega-\tfrac{\Omega_0}{2}).
\end{aligned}
\end{equation}

The time-dependent DOS for this system is given by:
\begin{equation}
\mathcal{N}_d(\omega,T)=-\frac{1}{\pi}\int \frac{d^dk}{(2\pi)^d} \text{Im}G^{\text{R}}(\textbf{k};\omega,T). 
\end{equation}
Using, the Lorentzian representation of the delta function we may write this as:
\begin{equation}
\begin{aligned}
\mathcal{N}_d(\omega,T)&=\int \frac{d^dk}{(2\pi)^d} \delta(\omega-E_\textbf{k}) \\
&+\frac{2U_0}{\Omega_0}\cos(\Omega_0 T)\int \frac{d^dk}{(2\pi)^d} \delta(\omega-\tfrac{\Omega_0}{2}-E_\textbf{k}) \\
&-\frac{2U_0}{\Omega_0}\cos(\Omega_0 T)\int \frac{d^dk}{(2\pi)^d}\delta(\omega+\tfrac{\Omega_0}{2}-E_\textbf{k}). 
\end{aligned}
\label{aeq:dos_t}
\end{equation}
Noting that each of these integrals has the same form as the undriven DOS, we can rewrite Eq (\ref{aeq:dos_t}) as:
\begin{widetext}
\begin{equation}
\mathcal{N}_d(\omega,T)=\mathcal{N}^{(0)}_d(\omega)+\frac{2U_0}{\Omega_0}\cos(\Omega_0 T)\left[ \mathcal{N}^{(0)}_d(\omega-\tfrac{\Omega_0}{2})-\mathcal{N}^{(0)}_d(\omega+\tfrac{\Omega_0}{2}) \right]
\label{aeq:dos_final}
\end{equation}
\end{widetext}
where $\mathcal{N}^{(0)}_d(\omega)$ is the DOS in $d$-dimensions associated with the dispersion $E_\textbf{k}$.

Consider the special case of $d=2$ and $E_\textbf{k}=\tfrac{\hbar^2 k^2}{2m}-\mu$. In this case, $\mathcal{N}^{(0)}_2(\omega)$ is a constant function of $\omega$, therefore Eq (\ref{aeq:dos_final}) is constant in $T$ and unchanged to linear order in $U_0$. This provides insight into why we observe little change in the magnitude of the 2D DOS shown in Fig~\ref{dos_plots}(a) and (b), the contributions from the Floquet copies cancel at linear order. 

Now, consider the case of $d=3$ and $E_\textbf{k}=\tfrac{\hbar^2 k^2}{2m}-\mu$. In this case $\mathcal{N}^{(0)}_3(\omega)\propto\sqrt{\omega}$, therefore, unlike the 2D case, the corrections do not cancel. Instead the linear-order corrections provide a net suppression at $T=0$ since, for $\tfrac{\Omega_0}{2}<\omega$, $\mathcal{N}^{(0)}_3(\omega+\tfrac{\Omega_0}{2})>\mathcal{N}^{(0)}_3(\omega-\tfrac{\Omega_0}{2})$. However, this suppression will disappear at $T=\pi/2\Omega_0$ due to the vanishing of cosine at this point in the period. Furthermore, this suppression will turn into an enhancement when the cosine is negative. This explains why the 3D DOS in Figs~\ref{dos_plots}(c) and (d) is slightly suppressed at $T=0$ but unchanged at $T=\pi/2\Omega_0$.

\bibliographystyle{apsrev}
\bibliography{Odd_Frequency}

\end{document}